\documentclass[aps,pra,twocolumn,showpacs,superscriptaddress,10pt]{revtex4-1}

\usepackage{amsmath}
\usepackage{amsfonts}
\usepackage{amssymb}
\usepackage{graphicx}
\usepackage{color}
\usepackage{xcolor}

\newcommand{\dd}{\mathrm{d}}
\newcommand{\ee}{\mathrm{e}}
\newcommand{\xx}{\mathbf{x}}
\newcommand{\rr}{\mathbf{r}}

\newcommand{\qq}{\mathbf{q}}
\newcommand{\pp}{\mathbf{p}}
\newcommand{\vecphi}{\mathbf{\Phi}}

\usepackage{tikz}
\usetikzlibrary{positioning,arrows}
\usetikzlibrary{decorations.pathmorphing}
\usetikzlibrary{decorations.markings}
\usetikzlibrary{calc}
\usetikzlibrary{patterns}
\usetikzlibrary{shapes.misc}
\tikzset{
phys/.style={thick, postaction={decorate}, decoration={markings, mark=at position .7 with {\arrow[]{triangle 45}}}},
res/.style={thick, decorate, decoration={snake,segment length=7, amplitude=1.5}},
arrow/.style={thick, draw=white, postaction={decorate}, decoration={markings, mark=at position .6 with {\arrow[black]{triangle 45}}}}
 }

\begin{document}
\title{Dynamical crossovers in prethermal critical states}
\author{Alessio Chiocchetta}
\affiliation{Institut f\"{u}r Theoretische Physik, Universit\"{a}t zu K\"{o}ln, D-50937 Cologne, Germany}
\affiliation{SISSA --- International School for Advanced Studies and INFN, via Bonomea 265, I-34136 Trieste, Italy}
\author{Andrea Gambassi}
\affiliation{SISSA --- International School for Advanced Studies and INFN, via Bonomea 265, I-34136 Trieste, Italy}
\author{Sebastian Diehl}
\affiliation{Institut f\"{u}r Theoretische Physik, Universit\"{a}t zu K\"{o}ln, D-50937 Cologne, Germany}
\author{Jamir Marino}
\affiliation{Institut f\"{u}r Theoretische Physik, Universit\"{a}t zu K\"{o}ln, D-50937 Cologne, Germany}

\begin{abstract}
We study the prethermal dynamics of an interacting quantum field theory with a $N$-component order parameter and $O(N)$ symmetry, suddenly quenched in the vicinity of a dynamical critical point. {Depending on the initial conditions}, the evolution of the order parameter, and of the response and correlation functions, {can} exhibit a temporal crossover between universal dynamical scaling regimes governed, respectively, by a quantum and a classical prethermal fixed point, as well as a crossover from a Gaussian to a non-Gaussian prethermal dynamical scaling. Together with a recent experiment, this suggests that quenches may be used in order to explore the rich variety of dynamical critical points occurring in the non-equilibrium dynamics of a quantum many-body system. We illustrate this fact by using a combination of  renormalization group techniques and a non-perturbative large-$N$ limit.
\end{abstract}


\date{\today}
\maketitle

\emph{Introduction ---}
An isolated many-body system, suddenly brought out of equilibrium, can linger over a non-thermal quasi-steady state before thermalization occurs at later times triggered by inelastic scattering processes. This phenomenon, known as prethermalization, is ubiquitous in physics: first predicted in the context of heavy-ion collisions~\cite{Berges2004a} and early-universe inflationary dynamics~\cite{Aarts2000}, it has been only recently recognized in condensed matter and cold-atoms physics~\cite{Gring2012,*Langen2013,*Langen2015,*Langen2016,*Moeckel2008,*Marino2012bis,*Marinolong2012,*Stark2013,*Mitra2013,*Marcuzzi2013,*Bertini2015,*Demler2015,*Marcuzzi2016, *Tavora2013,*Buchhold2016,*Kollath2007,*Berges2007,*Berges2008,*Rosch2008,*Scheppach2010,*Barnett2011,*Schole2012,*Poletti2012,*Nowak2012,*Poletti2013,*Nessi2014,*Kastner2015,*Nowak2014,*Orioli2015,*Karl2016}.

Prethermal states can host a novel  class of critical phenomena, since the  gap associated with these non-equilibrium (NEQ) states can vanish upon choosing properly the initial conditions and parameters of the system. Quenching a quantum many-particle system above or below the corresponding critical points results in qualitatively different NEQ dynamics~\cite{Barankov2004,*Barankov2006,*Yuzbashyan2006,*Gurarie2009,*Eckstein2009,*Sciolla2010,*Schiro2010,*Sciolla2011,*Schiro2011,*Foster2013,*Tsuji2013, *Peronaci2015,*Yuzbashyan2015, *Zunkovic2016,Gambassi2011}. For instance, in the case of an interacting bosonic quantum field, quenching above or below its dynamical critical point induces respectively an exponential relaxation of local observables or a coarsening dynamics of correlation functions~\cite{Sciolla2013,Chandran2013,Maraga2015}.
By quenching, instead, close to the dynamical critical point, correlation functions display non-equilibrium dynamical scaling~\cite{Gambassi2011, Chiocchetta2015, Maraga2015, Chiocchetta2016} similar to critical aging in classical dissipative systems~\cite{Janssen1989, Calabrese2005} with a consequent delay of thermalization and stabilization of the associated prethermal states. These NEQ dynamical critical points can both be  of quantum and classical nature, depending on the strength of the quantum quench performed at initial times~\cite{Chiocchetta2015, Chiocchetta2016}. 

Interestingly, a recent experiment~\cite{Oberthaler2015} has observed the emergence of dynamical scaling in a one-dimensional two-component Bose gas quenched close to criticality. The correlation length of the order parameter displayed there a dynamical crossover between a short-time critical  and a long-time non-critical regime. This suggests the possibility to explore the variety of the dynamical critical points occurring in the prethermal dynamics of a quantum many-body system via quantum quenches, which expose the temporal crossover between different  scaling behaviours.

In this work, we propose a minimal model in which it is possible to observe and control the real-time crossovers affecting the dynamical critical scaling of the order parameter and of the two-times correlation and response functions; these crossovers are determined by a broader class of NEQ critical points compared to those of the experiment mentioned above~\cite{Oberthaler2015}.
We provide numerical evidence of  this fact in an interacting bosonic field theory with $N$ components and $O(N)$ symmetry in the exactly solvable limit $N\to\infty$, and we interpret our findings via a renormalization-group (RG) analysis. The key results of our analysis are: \emph{(i) The existence of two temporal crossovers,} one involving the quantum and classical prethermal fixed points (FPs) associated with shallow and deep quenches, respectively~\cite{Chiocchetta2015, Chiocchetta2016}, and the other characterizing the transition from Gaussian to non-Gaussian prethermal dynamical scaling. While the first one occurs at the inverse of an analogue of the de Broglie thermal momentum scale, with the pre-quench value of the inverse of the correlation length replacing the temperature, the second scale sets in at the inverse of the quartic interaction strength. \emph{(ii) Crossover through RG:} These two scales are identified deriving flow equations which encompass the scaling regimes controlled by the quantum and the classical prethermal FPs, respectively. 

\emph{Dynamical  transitions---} 
Let us consider  a $N$-component bosonic order parameter $\vecphi=(\phi_1, \phi_2, ..., \phi_N)$ in $d$ spatial dimensions, with an $O(N)$-symmetric Hamiltonian 
\begin{equation}
\label{eq:hamiltonian}
H(c,r,u) = \!\!\int_\xx \left[\frac{1}{2}\mathbf{\Pi}^2 + \frac{1}{2}(c\nabla\vecphi)^2 + \frac{r}{2}\vecphi^2+\frac{u}{4!N}(\vecphi^2)^2\right],
\end{equation}
with $\int_\xx \equiv \int \dd^d x$, $\mathbf{\Pi}$ the $N$-component momentum canonically conjugate to $\vecphi$, $c > 0$ the speed of the quasi-particles, $r$ parametrising the distance from the critical point  and $u>0$ the strength of the leading non-linearity. The model requires the presence of an ultraviolet cutoff $\Lambda$, which is physically the inverse of a microscopic length scale, e.g., the lattice spacing.
The Hamiltonian~\eqref{eq:hamiltonian} finds a wide range of applications in particle physics and cosmology~\cite{Berges2004b}, and at equilibrium it belongs to the same universality class as several condensed matter systems~\cite{Sachdevbook, *Sondhi, *Vojta}, such as the Ising ($N = 1$) and Heisenberg ($N = 3$) models, and the Bose-Hubbard model at the particle-hole symmetric point ($N=2$). In the limit $N\to\infty$, the model becomes exactly solvable, since diagrammatic corrections beyond one loop are parametrically small in $1/N$, which allows a self-consistent closure of the hierarchy of correlation functions~\cite{ZinnJustinbook}. Remarkably, the large-$N$ limit captures several qualitative features of the equilibrium phase diagram of the $O(N)$ model for finite $N$ and $d$~\cite{Sachdevbook}. 

We assume that the system is prepared in the ground state of the non-interacting Hamiltonian $H(c_0,r_0,0)$ (with $r_0\equiv \Omega_0^2$), and then quenched at $t=0$  as $H(c_0,r_0,0) \mapsto H(c,r,u)$, with $c=1$ for simplicity. Since thermalizing inelastic collisions are expected to be effective only after times $\sim N/u^2$~\cite{Berges2004b, Sotiriadis2010}, the prethermal state actually becomes the  asymptotic steady state of the dynamics for $N\to\infty$~\cite{Sciolla2013,Chiocchetta2015,Smacchia2015,Maraga2015}. 

In the prethermal stage of the dynamics, one finds from perturbation theory at the leading order~\cite{Sotiriadis2010,Chiocchetta2015,Chiocchetta2016} a time-dependent dressed value $r_{\text{eff}}(t)$ of $r$, representing the contribution of fluctuations to its Gaussian value
\begin{equation}
\label{eq:effmass}
r_{\text{eff}}(t) = r + u\frac{N+2}{6N} \int \frac{\dd^dp}{(2\pi)^d}~\langle \phi_{\pp}(t)\phi_{-\pp}(t)  \rangle,
\end{equation}
where $\phi_{\pp}$ denotes the Fourier transform of one of the components of $\mathbf{\Phi}$.
As a consequence of dephasing, $r_{\text{eff}}(t)$ has a well-defined limit $r_{\text{eff}}(\infty)\neq 0$ for large times. When $r$ is tuned to a critical value $r_c(c_0,\Omega_0,u)$,  $r_{\text{eff}}(\infty)$ vanishes and, consequently, the correlation length $\xi = r_{\text{eff}}(\infty)^{-1/2}$ diverges, thus signalling the onset of a dynamical transition~\cite{Sciolla2013,Chandran2013,Smacchia2015}. 
%
\begin{figure}[t!]
\includegraphics[width=7.5cm]{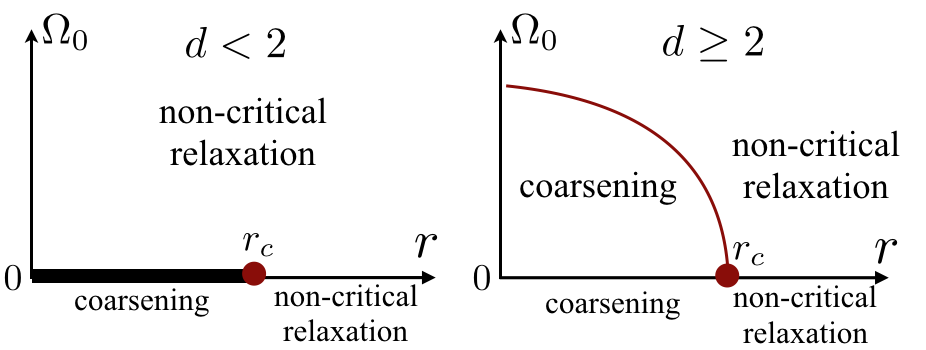}
\caption{(Color online). A dynamical transition separating a non-critical relaxation from  coarsening  occurs at $r=r_c(c_0,\Omega_0,u)$. For $d<2$ (left panel), a dynamical transition occurs only for $\Omega_0=0$, while any finite value of $\Omega_0$ prevents it;  for $d>2$, instead, the dynamical transition takes place also for finite values of $\Omega_0$.}
\label{fig:phase-diagram}
\end{figure}
%
Figure~\ref{fig:phase-diagram} briefly summarizes the phase diagram of the dynamical transition associated with the quench dynamics of model~\eqref{eq:hamiltonian}~\cite{Gambassi2011, Sciolla2013, Chandran2013, Chiocchetta2015, Maraga2015}.
For $r>r_c(c_0,\Omega_0,u)$, the system undergoes a non-critical relaxation characterized by a finite value of the correlation length and correlation time, and by a vanishing value of the order parameter. At $r = r_c(c_0,\Omega_0,u)$, the slow modes are characterized by aging dynamics, in which  the retarded $G_R(p,t,t')\equiv -i\langle[\phi_\pp(t),\phi_{-\pp}(t')]\rangle$ and Keldysh $G_K(p,t,t')\equiv-i\langle\{\phi_\pp(t),\phi_{-\pp}(t')\}\rangle$ functions acquire the scaling forms $G_R(p \to 0,t,t')\propto -t (t'/t)^{\theta}$ and $G_K(p\to 0,t,t')\propto (tt')^{2-2\theta}$ for $t\gg t'$, characterized by a  non-equilibrium universal exponent $\theta$ \cite{Chiocchetta2015, Maraga2015}. Moreover, when the system is prepared with a finite initial  value $M_0 \equiv M(t=0)$ of the order parameter $M(t) \equiv \langle \phi(t) \rangle$ and subsequently quenched to the critical point, $M(t)$  exhibits a remarkable universal non-monotonic dependence on time $M(t) = M_0t^\theta \mathcal{M}(M_0 t^{\theta + \beta/(\nu z)})$, with $\mathcal{M}(x) \approx x^{-1}$ for $x \gg 1$~\cite{Janssen1989, Calabrese2005} --- a further hallmark of aging dynamics. For $r<r_c(c_0,\Omega_0,u)$, instead, the system undergoes anomalous~\cite{Sciolla2013,Chandran2013,Maraga2015} \emph{coarsening}~\cite{Bray1994,Biroli2015,Cugliandolo2015}, and correlation functions are characterized by a  dynamical scaling, caused by the formation of growing domains with different values of the order parameter.

\emph{Equilibrium vs. prethermal criticality} --- The very nature of the prethermal critical points at $r=r_c(c_0,\Omega_0,u)$ strongly depends on the pre-quench parameter $\Omega_0$: if $\Omega_0=0$, the critical properties display \emph{quantum} features, while for $\Omega_0>0$ they are \emph{classical}, in the sense specified further below (see also  Fig.~\ref{fig:phase-diagram}). 
These quantum/classical prethermal critical points are analogous to the zero/high temperature ones of equilibrium phase transitions~\cite{Sachdevbook, *Sondhi, *Vojta}. 
This correspondence can be rationalized by highlighting two key properties of the quench problem. First, one can inspect the small-momentum scaling of the Gaussian ($u=0$)  $G_K$~\cite{Chiocchetta2015} at equal times and at the critical point  $r=0$: if $\Omega_0=0$, then  $G_K(p,t,t) \sim 1/p$, i.e. the dependence on $p$ is the same as in equilibrium at zero temperature;  for $\Omega_0\neq0$, instead,  $G_K(p,t,t) \sim \Omega_0/p^2$, as it happens in equilibrium at  finite temperature~\cite{Calabrese2006}. 
 Second, the integral on the r.h.s. of Eq.~\eqref{eq:effmass}  diverges as $r\to 0$ in $d < 3(4)$ for the  quantum (classical) critical point, signalling the breakdown of the validity of perturbation theory around the Gaussian FP and thus determining an upper critical dimensionality $d_u = 3(4)$~\cite{Smacchia2015,Chiocchetta2015}. Moreover, the same integral  diverges for any value of $r$ in $d< 1(2)$ for the quantum (classical) critical point, possibly indicating a lower critical dimensionality $d_l =1(2)$~\cite{Chiocchetta2016}. As a result, the values of $d_u$ and $d_l$ are the same as in the corresponding equilibrium theory~\cite{Sachdevbook, *Sondhi, *Vojta}. 

\begin{figure}[t!]
\includegraphics[width=6.5cm]{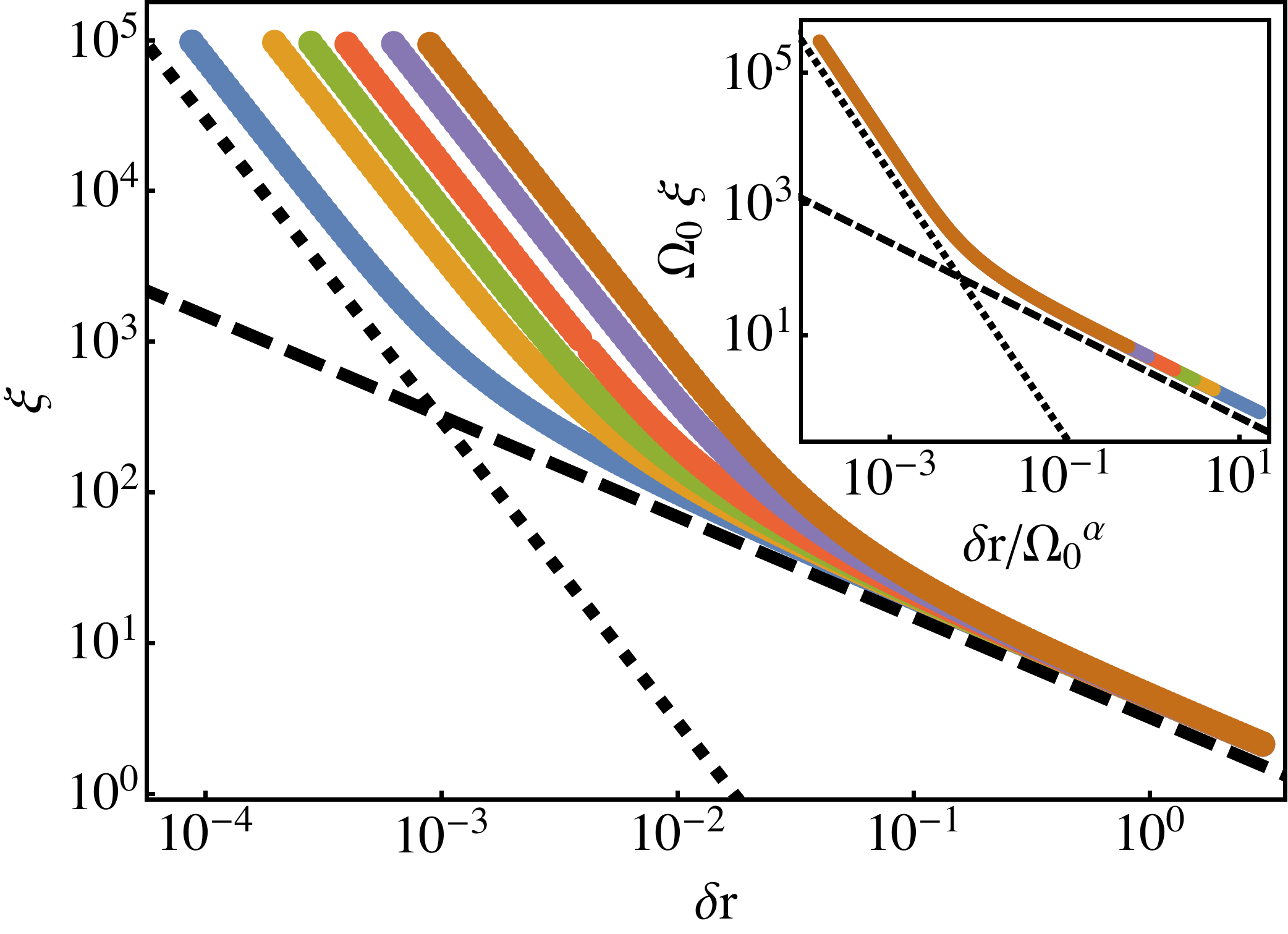} 
\caption{(Color online). Correlation length $\xi$ for the large-$N$ limit of Eq.~\eqref{eq:hamiltonian}, as a function of the distance from the critical point $\delta r \equiv |r-r_c|$. Results for $d=2.5$, $u=10$, $c_0 = 100$, and various values of $\Omega_0$ ($\Omega^2_0 = 0.1, 0.5,1,2,5,10$, from the lowermost to the uppermost curves, respectively). The dashed line is proportional to $r^{-3/2}$, while the dotted one is proportional to $r^{-2}$, corresponding to the quantum and classical scaling, respectively. 
Inset: $\Omega_0\xi $ vs. $\delta r/ \Omega_0^{\alpha}$, based on the same data as the main plot, with $\alpha \approx 1.5$ determined numerically.}
\label{fig:correlation-length}
\end{figure}
%
\begin{figure}[t!]
\includegraphics[width=7.5cm]{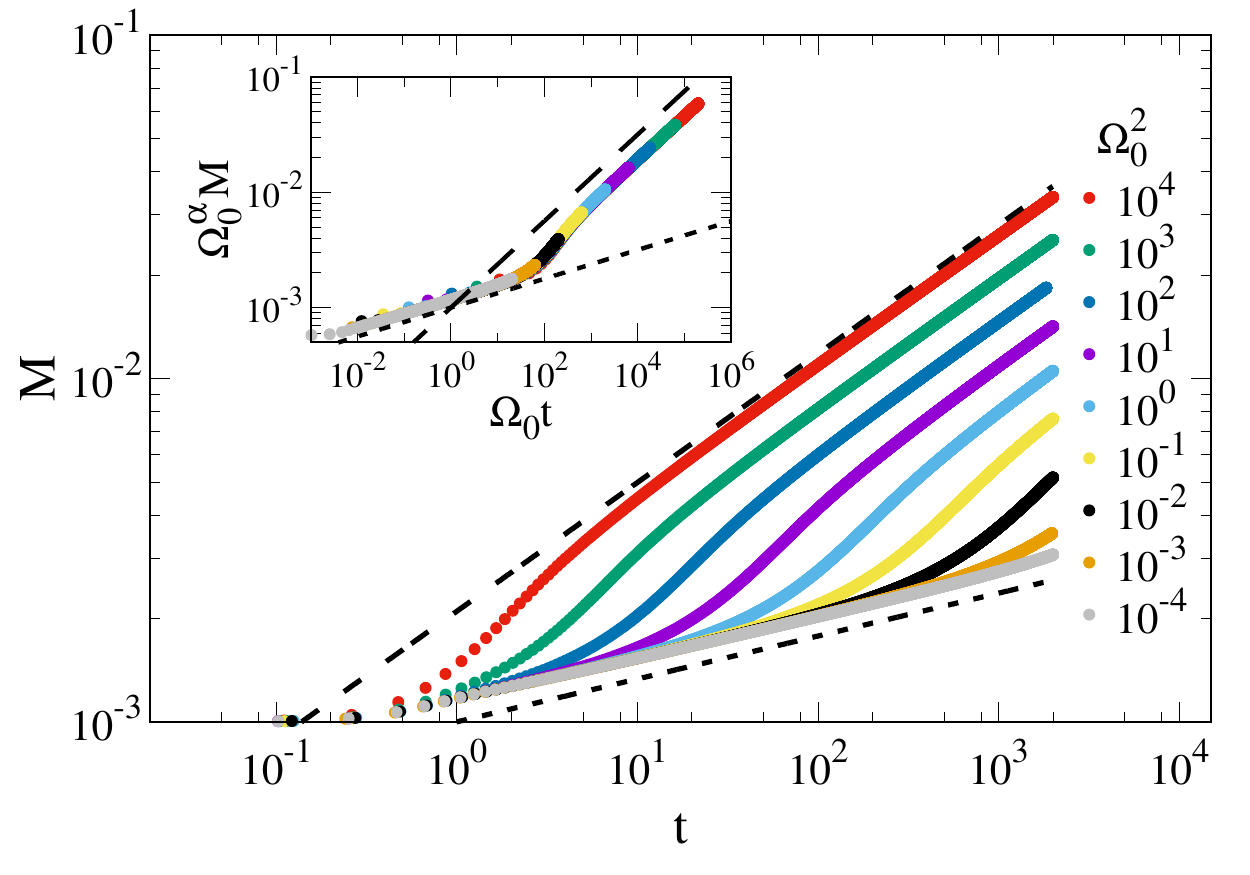} 
\caption{(Color online). Magnetization $M(t)$ for the large-$N$ limit of Eq.~\eqref{eq:hamiltonian}, as a function of time $t$ for $d=2.5$, $u=1$, $c_0 = 100$, $r = r_c$, and for several values of $\Omega_0$, reported in the legend of the main plot.
Inset: $\Omega_0^\alpha M $ vs. $\Omega_0 t$, based on the same data as the main plot, with $\alpha \approx 0.12$ determined numerically.}
\label{fig:magnetization}
\end{figure}
%
\emph{Large-$N$ limit} --- In order to demonstrate the existence of crossovers in the dynamics of observable quantities, we first study the large-$N$ limit of the Hamiltonian~\eqref{eq:hamiltonian},  which is exactly solvable. While previous works~\cite{Sciolla2013,Chiocchetta2015,Smacchia2015,Maraga2015, Chiocchetta2016} focussed  on the classical scaling, we provide here  evidence  for the emergence of an unnoticed quantum scaling behaviour. We first consider the correlation length $\xi \sim(\delta r)^{-\nu}$ in the stationary non-thermal state as a function of the distance $\delta r \equiv |r - r_c|$ from the critical point~\ref{app:app1}: in Fig.~\ref{fig:correlation-length}, we show $\xi$ for various values of $\Omega_0$. One observes that $\xi$ crosses over from the classical scaling $\xi \sim |\delta r|^{-1/(d-2)}$ for $\delta r \to 0$, to the quantum scaling $\xi \sim |\delta r|^{-1/(d-1)}$ for larger values of $\delta r$; the crossover occurs at $\delta r\sim\Omega^2_0$, and the different curves collapse onto a master curve after rescaling (inset).
A similar behaviour is exhibited as a function of time by the  order parameter $M(t)$, whose behaviour is expected to be $M(t) \sim t^\theta $ for $t \ll M_0^{-\nu z/(\beta + \nu z \theta)}$~\cite{Chiocchetta2016}. Figure~\ref{fig:magnetization} shows  $M(t)$  for various values of $\Omega_0$: each curve crosses over from the quantum behaviour $\propto t^{(3-d)/4}$ at early times, to the classical one $\propto t^{(4-d)/4}$ at late times: after a rescaling by $\Omega_0$ (see inset), all the curves collapse onto a master curve.   
These crossovers  are rationalized further below in terms of an RG picture.       

\emph{RG analysis} --- In order to study the properties of the prethermal critical points discussed above, we use a functional RG (FRG) scheme~\cite{Berges2002,Chiocchetta2016bis}. In this respect, it is convenient to consider the Keldysh action~\cite{Altlandbook2010,*Kamenevbook2011} associated with Eq.~\eqref{eq:hamiltonian}, written in terms of the $N$-components classical and quantum fields $\vecphi_c $ and $ {\vecphi}_q$, respectively:
\begin{multline}
\label{actionbulk}
S = \int_{\xx,t}  \bigg[ \dot{{\vecphi}}_q \cdot \dot{\vecphi}_c - c^2( \nabla {\vecphi}_q ) \cdot ( \nabla \vecphi_c ) - r {\vecphi}_q \cdot \vecphi_c \\
- \frac{u_c}{12N}  {\vecphi}_q \cdot \vecphi_c \vecphi_c^2 -\frac{u_q}{12N}  {\vecphi}_q \cdot \vecphi_c {\vecphi}^2_q \bigg],
\end{multline}
where $u_c$ and $u_q$ parametrize the so-called classical and the quantum vertices~\cite{Kamenevbook2011} respectively (their value coincides with $u$ in Eq.~\eqref{eq:hamiltonian} at the microscopic scale). 
We also emphasise that the RG approach is applicable for generic values of $N$.
In order to apply the  FRG scheme~\cite{Chiocchetta2016bis}  for studying the non-equilibrium universal dynamics generated by the Hamiltonian~\eqref{eq:hamiltonian}, it is necessary to encode the initial conditions into a Keldysh action defined on the time surface induced by the quench~\cite{Chiocchetta2015,Chiocchetta2016}. Such action involves the fields $\vecphi_{0,c}$ and ${\vecphi}_{0,q}$, corresponding to the initial values of the classical and quantum fields, respectively~\ref{app:app2}. In particular, the aging exponent $\theta$ is related to the anomalous dimension $\eta_0$ of the quantum boundary field ${\vecphi}_{0,q}$ as $\theta=\eta_0/z$~\cite{Janssen1989, Calabrese2005,Chiocchetta2015, Chiocchetta2016}, where the dynamical critical exponent is $z=1$ up to one loop (or to leading order in $1/N$)~\cite{Chiocchetta2015,Maraga2015}. 

As expected on the basis of the analogy  with equilibrium criticality discussed above, the one-particle irreducible effective action associated with $S$, and the related beta functions~\ref{app:app2} admit scaling solutions corresponding to both the quantum ($u_{c,q}\sim k^{3-d}$) and the classical ($u_c \sim k^{4-d}$, $u_q\sim k^{2-d}$) canonical power counting, where $k$ is a running momentum  scale introduced to regularize the FRG equation~\ref{app:app2}. The associated FPs are captured by expressing the beta functions~\ref{app:app2} in terms of the dimensionless parameters $\widetilde{r} \equiv r/k^2 $, $\widetilde{\Omega}_0 \equiv \Omega_0/(c_0 k)$, $\widetilde{u}_c \equiv u_c a_d c_0 k^{d-3}(1+\widetilde{\Omega}_0^2)^{1/2}/d$, and $\widetilde{u}_q \equiv u_q a_d k^{d-3}/[d (1+\widetilde{\Omega}_0^2)^{1/2}]$, where $a_d = 2/[(4\pi)^{d/2}\Gamma(d/2)]$ is a numerical factor. The  dimensionless RG flow equations (at one loop) are written in terms of the running parameter $\ell$ with $\ell=\log\left(\Lambda/k\right)$, which flows from $\ell=0$ at ultraviolet scales $k=\Lambda$ towards $\ell\to\infty$ at infrared scales $k\ll\Lambda$; they read:
\begin{subequations}
\label{eq:flows}
\begin{align}
\frac{\dd \widetilde{r}}{\dd \ell} &= 2\widetilde{r} + \frac{N+2}{12N}\frac{\widetilde{u}_c}{(1+\widetilde{r})^2}, \label{eq:RG-1}\\
\frac{\dd \widetilde{u}_{c,q}}{\dd \ell} & = -\widetilde{u}_{c,q}\left[D_{c,q}(\widetilde{\Omega}_0) + \frac{N+8}{6N}\frac{\widetilde{u}_c}{(1+\widetilde{r})^3}\right], \label{eq:RG-2}\\
\frac{\dd \widetilde{\Omega}_0}{\dd \ell} &=  \widetilde{\Omega}_0, \label{eq:RG-3}
\end{align}
\end{subequations}
while the anomalous dimension $\eta_0$ of the field $\bar{\vecphi}_{0,q}$ is given by $\eta_0 = - ( N + 2 ) \widetilde{u}_c/[24N(1+\widetilde{r})^3]$. In Eqs.~\eqref{eq:flows} we defined the running canonical dimensions $D_{c,q}(\widetilde{\Omega}_0)$ of $\widetilde{u}_{c,q}$ as
\begin{equation}
D_{c,q}(\widetilde{\Omega}_0)=d-3 \mp \frac{\widetilde{\Omega}_0^2}{1+\widetilde{\Omega}_0^2},
\end{equation}
which play a crucial role in describing the crossover between the quantum and the classical prethermal FPs.

We now show how the crossovers previously discussed on the basis of the exact solution of the model for $N\to\infty$ (see Figs.~\ref{fig:correlation-length} and~\ref{fig:magnetization}), emerge from the RG flow described by Eqs.~\eqref{eq:flows}.%

\emph{Crossover between quantum and classical Wilson-Fisher FPs ---} 
Equation~\eqref{eq:RG-3} admits two distinct FPs for $\widetilde{\Omega}_0$, i.e., $\widetilde{\Omega}_0^*=0$ and $\widetilde{\Omega}_0^*=+\infty$, corresponding to the quantum and classical prethermal critical points respectively, as we further detail below. For each value of $\widetilde{\Omega}_0^*$, Eqs.~\eqref{eq:RG-1} and~\eqref{eq:RG-2} possess a Gaussian and a Wilson-Fisher (WF) FP, which are stable for $D_{c,q}(\widetilde{\Omega}_0^*) >0$ and $D_{c,q}(\widetilde{\Omega}_0^*) < 0$, respectively. Note that $D_{c}(0) =D_{q}(0)= d-3$, while $D_c(+\infty) = d-4$ and $D_q(+\infty) = d-2$, thus recovering the correct upper critical dimensionality $d_u$ previously discussed (accordingly, $u_q$ becomes irrelevant in a RG sense for $d>2$ at the classical FP). 
Compared to the classical one, the quantum prethermal FP (with $\tilde{\Omega}_0^*=0$) has an  infrared unstable direction in addition to  the one related to $r$, since $\widetilde{\Omega}_0$ (equivalently, $\Omega_0$) acts as a relevant perturbation unless it is fine-tuned to zero, in the same way as temperature destabilizes equilibrium quantum critical points~\cite{Sondhi,Sachdevbook, Vojta}: accordingly, any finite value of $\Omega_0$ induces a crossover from the quantum to the classical FP, as  illustrated in the left panel of Fig.~\ref{fig:correlation-length}.
This crossover is accounted for in the RG Eqs.~\eqref{eq:flows} by the running dimensions $D_{c,q}(\widetilde{\Omega})$, which flow as a result of the RG equation of $\widetilde{\Omega}_0$~\eqref{eq:RG-3}. We emphasize that this crossover occurs only when both the FPs exist and are stable with respect to $r$, i.e., for $2<d<3$. For $d<2$ (i.e., below the lower critical dimension of the classical FP) a finite value of $\Omega_0$ will induce a crossover from the quantum FP towards a non-critical regime. These crossover phenomena are analogous to what is observed in the experiment reported in Ref.~\cite{Oberthaler2015}.
The microscopic value $\Omega_0$ in fact determines a momentum scale $k_0 \sim \Omega_0$ (corresponding to an RG scale $\ell_0 \sim \log(\Lambda/\Omega_0)$) which is analogous to the inverse of the thermal de Broglie length for quantum phase transitions, with $\Omega_0$ replacing the temperature~\cite{Sondhi,Vojta,Sachdevbook}. 

\emph{Crossover between Gaussian and Wilson-Fisher FPs ---}
Another  crossover occurring in the system  is the one involving Gaussian and WF FPs. First of all, Eqs.~\eqref{eq:flows} admit two different Gaussian FPs corresponding to $\widetilde{\Omega}^*_0=0$ and  $\widetilde{\Omega}^*_0=+\infty$ and given by $(\widetilde{r}^*,\widetilde{u}_{c,q}^*)=(0,0)$. Focusing below on the specific case in which $\widetilde{\Omega}_0$ takes one of its FP values $\widetilde{\Omega}_0^*$, for $D_c(\widetilde{\Omega}_0^*) < 0$ the Gaussian FP is unstable and any finite value of $\widetilde{r}$ and $\widetilde{u}_c$ will induce a crossover towards a WF critical point or to a non-critical point with $\widetilde{r}^*=+\infty$. This crossover occurs on a Ginzburg momentum scale $k_G$~\cite{Pelissetto2002,Sachdevbook}, which is proportional to the microscopic value of the interaction $u$. The scale $k_G$   signals also the breakdown of perturbation theory around the Gaussian FP; a simple calculation along the lines of Ref.~\cite{Marino2016bis} shows that $k_G \sim (u c_0)^{1/(3-d)}$ for the quantum scaling limit, while $k_G \sim (u \Omega_0 )^{1/(4-d)}$ for the classical case.
When $\widetilde{\Omega}_0$ is not at its FP value, an interplay between the quantum-classical and Gaussian-WF crossovers occurs, and the scenario enriches, depending on  $d$ and on the relative values of the length scales $k_0$ and $k_G$. 

\emph{Dynamical crossovers ---} The crossovers described above can be observed in the scaling behaviour of static and, especially, dynamic quantities.   
For instance, the correlation length $\xi \sim|\delta r|^{-\nu}$ may cross over between different algebraic behaviours characterized by different values of the exponent $\nu$: for the Gaussian FP $\nu = 1/2$, while $\nu = 1/2 + \epsilon (N+2)/[4(N+8)] + O(\epsilon^2)$ for the quantum ($\epsilon = 3-d$) and classical ($\epsilon = 4-d$) FPs~\cite{Smacchia2015,Chiocchetta2015,Chiocchetta2016}. These crossovers are controlled by the ratios $\delta r/k_0^2$ and $\delta r/k_G^2$, in agreement with the result obtained for $N\to\infty$. 

Similarly, the Keldysh and retarded functions $G_{K,R}$, as well as the  order parameter $M(t)$  discussed after Eq.~\eqref{eq:effmass}, may cross over between different algebraic time dependence, i.e., $M(t)\propto t^\theta$,  with $\theta =0$ for the Gaussian FP, and $\theta = \epsilon (N+2)/[4(N+8)]$ for the quantum ($\epsilon = 3-d$), and classical ($\epsilon = 4-d$) FPs~\cite{Chiocchetta2015,Chiocchetta2016} as shown in Fig.~\ref{fig:magnetization} for $N\to\infty$. The dynamical crossovers involving the quantum and classical WF FPs, as well as the one involving the Gaussian and WF FPs, are controlled respectively  by the dimensionless quantities $k_0 t$ and $k_G t$, {i.e., they occur respectively at the two characteristic times,  $t_0\sim 1/k_0$ and  $t_G\sim1/k_G$}.

{ \emph{Onset of thermalization} --- 
The inelastic collisions between quasiparticles arising at two-loops and responsible for thermalization are signalled by the generation of dissipative terms in Eq.~\eqref{actionbulk}~\cite{Mitra2011, Chiocchetta2016}. 
We estimate the thermalization time $t_\text{th}$ from the RG scale at which these terms become sizeable, obtaining~\ref{app:app3}, at large $N$, $t_{th}\sim N/(\Lambda(\widetilde{u}^*_c)^2)$ with  $\widetilde{u}^*_{c}$ the pre-thermal FP value of $\widetilde{u}_c$. Accordingly, one can make the time scales $t_G$ and $t_0$, determining the prethermal scaling regimes, much smaller than $t_\text{th}$ by tuning the parameters $u$, $\Omega_0$ and $N$.}
%
%

\emph{Conclusions---} In this work we  demonstrated the emergence of crossovers between different  scaling regimes associated with dynamical transitions in a $d$ dimensional quantum many-body system  with a $O(N)$-symmetric order parameter, using a FRG approach and  an exact  solution for $N\to\infty$. 
These crossovers are displayed by the evolution of observables and correlation functions which are within experimental reach~\cite{Cheneau2012,Trot2011,Meinert13}. 
The model considered here is a generalization of the one describing the  experiment reported in Ref.~\cite{Oberthaler2015}, where a dynamical crossover between different scaling behaviours was observed in a one-dimensional two-component Bose gas, and therefore our predictions can  be tested by performing  such experiment in two and three spatial dimensions. 
As an outlook, our ideas may be extended to study the dynamical phase diagram of open~\cite{Gagel2014,*Gagel2015} and driven-dissipative~\cite{Sieberer2013,*Marino2016} quantum systems.

\emph{Acknowledgements ---}  We thank A. Rosch for comments on the manuscript. We acknowledge useful discussions with G. Biroli, L.~F. Cugliandolo, L. Foini, K. Hazzard, C. Kollath, J. Schmiedmayer. A.~C. acknowledges hospitality from the University of Cologne. S.~D. acknowledges funding by the German Research Foundation (DFG) through the Institutional Strategy of the University of Cologne within the German Excellence Initiative (ZUK 81), and by the European Research Council via ERC Grant Agreement n. 647434 (DOQS). J.~M. acknowledges support from the Alexander von Humboldt foundation.

\appendix

\begin{widetext}

\section{Large-$N$ limit}
\label{app:app1}
The dynamical phase transition can be conveniently studied in the large-$N$ limit of the Hamiltonian in Eq.~(1) in the main text, since in this limit the dynamical problem becomes exactly solvable, as discussed in detail in Refs.~\cite{Sciolla2013,Chandran2013,Smacchia2015,Maraga2015,Chiocchetta2016}. In fact, the limit $N\to \infty $ makes the self-consistent approximation in which the ${(\vecphi^2)}^2$ interaction term is replaced by a quadratic ``mean-field'' term $\propto \vecphi^2 \langle \vecphi^2\rangle$ exact.

In order to study the dynamics of the order parameter, we will assume that the $O(N)$ symmetry is broken in the initial state, such that the expectation value $\langle \vecphi(t=0) \rangle \neq 0$. The equations determining the time evolution of the system are given by~\cite{Sciolla2013, Chiocchetta2016}:
\begin{subequations}
\label{eq:HFNfin}
\begin{align}
\label{eq:HFNfin-1}
& \left[ \partial _t^2 + r(t) - \frac{u}{3}M^2(t) \right]M(t)= 0,\\
& \left[ \partial _t^2 + q^2 + r(t)  - \frac{u}{6N} iG_K^\parallel (\rr = 0,t,t) \right]G_K^\parallel (q,t,t') = 0,\label{eq:HFNfin-2}\\
& \left[ \partial _t^2 + q^2 + r(t) - \frac{u}{3}M^2(t) \right]G_K^\bot (q,t,t') = 0,\label{eq:HFNfin-3}\\
& r(t)= r + \frac{u}{2}\left[ M^2(t) + \frac{1}{2N}iG_K^\parallel (\rr = 0,t,t)  + \frac{N - 1}{6N}iG_K^ \bot (\rr = 0,t,t) \right],\label{eq:HFNfin-4}
\end{align}
\end{subequations}
where $M(t) = \langle \phi_\alpha \rangle$ and $\alpha$ is the component along which the symmetry is broken; $G_K^{\|}$ and $G_K^{\bot}$ are the Keldysh Green's functions associated with the longitudinal and with a generic transverse component of the field, respectively, while the function $r(t)$ is a time-dependent effective parameter, self-consistently determined from Eq.~\eqref{eq:HFNfin-4}.

To solve the system of equations~\eqref{eq:HFNfin} one needs to specify the initial conditions for the average order parameter $M(t)$ and the Green's functions $G_K^{\|,\bot}$. As discussed in Refs.~\cite{Sciolla2013,Chiocchetta2016}, the suitable initial conditions are given by 
\begin{subequations}
\begin{align}
&M(t=0) = M_0, \\
&\partial_t M(t=0) = 0, \\
&G_K^{\parallel , \bot }(q,t = 0,t' = 0) = \left[ q^2 + r_{\parallel,\bot} (0)\right]^{-1/2},\\
&{\partial _t}G_K^{\parallel , \bot }(q,t = 0,t' = 0) = 0,\\
&{\partial _t}{\partial _{t'}}G_K^{\parallel , \bot }(q,t = 0,t' = 0) = \left[q^2 + r_{\parallel,\bot}(0)\right]^{1/2},
\end{align}
\end{subequations}
where $M_0$ is the initial value of the order parameter and where we have introduced the quantities
\begin{subequations}
\begin{align}
&r_\parallel(0) = r(0),\\
&r_ \bot(0) = r(0) - \frac{u}{3}M_0^2.
\end{align}
\end{subequations}

\section{Boundary functional renormalization group for a quantum quench}
\label{app:app2}
In order to apply the FRG scheme~\cite{Berges2002} to the case of a quantum quench discussed in the main text, it is necessary to supplement the action $S$ (Eq.~(3) in the main text) with a cutoff function $R_k(q)=K(q^2-k^2)\theta(k^2-q^2)$, which depends on the momentum scale $k$. More precisely, the function $R_k(q)$ is introduced as a quadratic term in the modified action $S_k \equiv S + \Delta S_k$, where $\Delta S_k =-\int_{t,\qq} R_k(q) \vecphi_q(-\qq) \vecphi_c(\qq)$. The effect of the cutoff fuction is to provide an effective finite correlation length to the otherwise critical modes, thus preventing the occurrence of infrared divergences in loop integrals~\cite{Berges2002,Litim2000}.

As a second step, one considers the $k$-dependent one-particle irreducible effective action $\Gamma_k$ (derived as the Legendre transform of the generating function associated with $S_k$), which can be also regarded as an effective action which has been coarse-grained on a volume $ k^{-d}$~\cite{Berges2002}. The effective action $\Gamma_k$ thus satisfies the FRG equation~\cite{Berges2002}
\begin{equation}
\label{eq:Wetterich}
\partial_k\Gamma_k=\frac{i}{2} \operatorname{Tr} \left[(\Gamma_k^{(2)}+R_k)^{-1}\partial_kR_k\right],
\end{equation}
where $\Gamma_k^{(2)}$ is the second functional derivative of the effective action with respect to classical and quantum fields. The label $k$ indicates that $\Gamma_k$ is a function of the running scale $k$ through the regulator $R_k$. In the following we will drop this label in order to simplify the notation.

In order to solve Eq.~\eqref{eq:Wetterich}, it is necessary to provide an ansatz for the form of the effective action $\Gamma$. Following Ref.~\cite{Chiocchetta2016bis}, we consider the following expression:
\begin{equation}
\label{eq:ansatz}
\Gamma = \Gamma_0 + \int_{\rr,t}\vartheta(t) \left[\dot{\vecphi}_q\cdot\dot{\vecphi}_c - (\nabla \vecphi)_q\cdot(\nabla \vecphi)_c - r\vecphi_q\cdot\vecphi_c -\frac{2 u_c}{4!N}\vecphi_q\cdot\vecphi_c \vecphi_c^2 -\frac{2 u_q}{4!N}\vecphi_q\cdot\vecphi_c \vecphi_q^2\right], 
\end{equation}
where $\vartheta(t)$ is the Heaviside step function while the boundary action $\Gamma_0$ contains information about the initial state and it is given by:
\begin{equation}
\label{eq:boundary-action}
\Gamma_0 = - \frac{1}{2} \int_\pp \left[ (c^2_0p^2+\Omega^2_{0q} )^{1/2} Z_0^2\vecphi^2_{0q}
 -\frac{1}{(c^2_0p^2+\Omega^2_{0c})^{1/2}} Z^2_0\dot{\vecphi}^2_{0q} + Z_0\vecphi_{0q}\dot{\vecphi}_{0c}+Z_0\dot{\vecphi}_{0q}\vecphi_{0c}\right].
\end{equation}
The quantum and classical pre-quench inverse correlation length~\cite{Chiocchetta2016} $\Omega_{0q,0c}$ occur as boundary couplings of the two boundary operators $\vecphi^2_{0q}$ and $\dot{\vecphi}^2_{0q}$, respectively: while they are identical at microscopic level, they may in principle flow differently under the RG transformation. In particular, the operator $\dot{\vecphi}^2_{0q}$ is irrelevant in the  RG sense, and therefore we will neglect it in the following analysis: this amounts to set $1/\Omega_{0c}$ to $0$, thus eliminating $\Omega_{0c}$ from the equations. Accordingly, below we will set $\Omega_0 \equiv \Omega_{0q}$ in order to simplify the notation. 
Further operators in the boundary action~\eqref{eq:boundary-action} are marginal or irrelevant in a RG sense for the instances of criticality considered in the main text. In particular, quartic couplings $u_{0c,0q}$ on the boundary, which would correspond to the presence of interactions in the initial state, scale respectively as $u_{0c,0q}\sim k^{2-d}$ and $u_{0c}\sim k^{3-d}$, $u_{0q}\sim k^{1-d}$ (in terms of the momentum scale $k$) for the quantum and classical canonical power counting considered in the following. These operators are therefore irrelevant in spatial dimensions close to the proper upper critical dimension of the quantum and classical FPs, and therefore they can be neglected in the following analysis, implying that our results hold for a broad class of initial NEQ conditions.

The derivation of $\beta$-functions from Eq.~\eqref{eq:Wetterich} is obtained in the symmetric phase, i.e., by setting to zero the background expectation value of the classical field. More precisely, we recall that the FRG equation~\eqref{eq:Wetterich} can be rewritten as~\cite{Chiocchetta2016bis}
\begin{equation}
\label{eq:FRG-decomposed}
\frac{\dd \Gamma}{\dd k} = \sum_{n=1}^{+\infty} \Delta \Gamma_n,
\end{equation}
where 
\begin{equation}
\Delta \Gamma_n = -\frac{i}{2} \sum_{\alpha,\alpha_1,\dots,\alpha_{2n}}\int_{x,y_1,\dots,y_n} \text{tr}\left[ G^0_{\alpha\alpha_1}(x,y_1)V_{\alpha_1\alpha_2}(y_1)G^0_{\alpha_2\alpha_3}(y_1,y_2) \dots V_{\alpha_{2n-1}\alpha_{2n}}(y_n)G^0_{\alpha_{2n}\alpha}(y_n,x)\frac{\dd R}{\dd k} \sigma\right],
\end{equation}
where we used the shorthand $x \equiv (\rr,t)$, while the matrix $G^0_{\alpha \beta}(x,x')$ is defined as
\begin{equation}
G^0_{\alpha \beta}(x,x') = \delta_{\alpha \beta} G^0(x,x') \equiv \delta_{\alpha \beta}
\begin{pmatrix} 
G_K(\rr-\rr',t,t') & G_R(\rr-\rr',t,t') \\
G_R(\rr-\rr',t',t) & 0
\end{pmatrix},
\end{equation}
where $G_{R,K}(\rr,t,t') = \int_\qq \ee^{i\qq \cdot \rr} G_{R,K}(q,t,t')$, and
\begin{equation}
G_R(q,t,t') = - \vartheta(t-t')\frac{\sin[\omega_q(t-t')]}{\omega_q}, \qquad iG_K(q,t,t') = \frac{\omega_{0q}}{\omega^2_q}\sin(\omega_q t)\sin(\omega_q t').
\end{equation}
In the equations above we introduced the pre- and post-quench dispersion relations $\omega^2_q = q^2 + r + (k^2-q^2)\vartheta(k^2-q^2)$ and $\omega^2_{0q} = c_0^2q^2 + \Omega^2_0 + c^2_0(k^2-q^2)\vartheta(k^2-q^2)$ modified by the regulator $R_k$, and  the matrix $V_{\alpha\beta}(x)$ as
\begin{align}
V_{\alpha\beta}(x) 
& = \vartheta(t)\frac{u_c}{6N}
\begin{pmatrix}
 \phi_c^{(\alpha)}\phi_q^{(\beta)} +\phi_c^{(\beta)}\phi_q^{(\alpha)} + \delta_{\alpha\beta}\vecphi_q\cdot\vecphi_c &  \delta_{\alpha\beta}\vecphi_c^2/2 +\phi_c^{(\alpha)}\phi_c^{(\beta)}\\
\delta_{\alpha\beta}\vecphi_c^2/2 + \phi_c^{(\alpha)}\phi_c^{(\beta)} & 0
\end{pmatrix} \nonumber \\
&  +\vartheta(t)\frac{u_q}{6N}
\begin{pmatrix}
 0 &  \delta_{\alpha\beta}\vecphi_q^2/2 +\phi_q^{(\alpha)}\phi_q^{(\beta)}\\
\delta_{\alpha\beta}\vecphi_q^2/2 +\phi_q^{(\alpha)}\phi_q^{(\beta)} & \phi_c^{(\alpha)}\phi_q^{(\beta)} +\phi_c^{(\beta)}\phi_q^{(\alpha)} + \delta_{\alpha\beta}\vecphi_q\cdot\vecphi_c
\end{pmatrix}.
\end{align}
The term $\Delta \Gamma_1$ in Eq.~\eqref{eq:FRG-decomposed} contains the renormalization of $r$ and $Z_0$, and it can be evaluated as follows:
\begin{align}
\label{eq:DGamma1}
\Delta \Gamma_1 
& = -\frac{i}{2}\sum_{\alpha\beta \gamma}\int_{\rr,\rr'}\int_0^{+\infty}\dd t\,\dd t' \, \text{tr}\left[ G^0_{\alpha\beta}(\rr-\rr',t,t')V_{\beta\gamma}(\rr',t')G^0_{\gamma \alpha}(\rr'-\rr,t',t) \frac{\dd R}{\dd k}\sigma\right] \nonumber \\
& = -\frac{i}{2}\sum_{\alpha}\int_{\rr,\rr'}\int_0^{+\infty}\dd t\,\dd t'  \,\text{tr}\left[ G^0(\rr-\rr',t,t')V_{\alpha\alpha}(\rr',t')G^0(\rr'-\rr,t',t) \frac{\dd R}{\dd k}\sigma\right] \nonumber \\
& = -\frac{i}{2}\sum_{\alpha}\int_{\rr,\qq}\int_0^{+\infty}\dd t\,\dd t' \, \text{tr}\left[ G^0(q,t,t')V_{\alpha\alpha}(\rr,t')G^0(q,t',t) \frac{\dd R}{\dd k}\sigma\right] \nonumber \\
& = -i k^{d+1}\frac{a_d}{d}\sum_{\alpha}\int_{\rr}\int_0^{+\infty}\dd t\,\dd t' \text{tr}\left[ G^0(k,t,t')V_{\alpha\alpha}(\rr,t')G^0(k,t',t) \sigma\right] \nonumber \\
& = -i k^{d+1}\frac{a_d}{d}\frac{u_c}{3N}\sum_{\alpha}\int_{\rr}\int_0^{+\infty}\dd t\,\dd t' \left\{ G_K(k,t,t')G_R(k,t',t) \left[ \phi_c^{(\alpha)}\phi_q^{(\alpha)} +\phi_c^{(\alpha)}\phi_q^{(\alpha)} + \delta_{\alpha\alpha}\vecphi_q\cdot\vecphi_c  \right]\right\} \nonumber \\
& = -i k^{d+1}\frac{a_d}{d}\frac{u_c(N+2)}{3N}\int_{\rr}\int_0^{+\infty}\dd t\,\dd t' \left[ G_K(k,t,t')G_R(k,t',t) \vecphi_q\cdot\vecphi_c\right] \nonumber \\
& =  k^{d+1}\frac{a_d}{d}\frac{u_c(N+2)}{12N}\int_\rr\int_0^{+\infty}\dd t\, \vecphi_q\cdot\vecphi_c \left[  \frac{\omega_{0k}}{\omega_k^4} + f_r(t) \right],
\end{align}
where
\begin{equation}
f_r(t) = - \frac{\omega_{0k}}{\omega_k^4}\left[ \cos(2\omega_kt) +\omega_k t \sin(2\omega_k t) \right].
\end{equation}
The function $f_r(t)$ averages to zero and therefore it is not expected to renormalize the post-quench parameter $r$, which is instead renormalized by the constant term in the integrand in the last line of Eq.~\eqref{eq:DGamma1}. On the other hand, $f_r(t)$ generates an infinite number of terms in the boundary action~\cite{Chiocchetta2016bis}: most of these terms are  irrelevant in the RG sense, except for that one which renormalizes the wave-function renormalization coefficient $Z_0$ of the boundary field $\vecphi_{0q}$.
The flow equations for $r$ and $Z_0$ read: 
\begin{subequations}
\label{eq:RG-eq-1}
\begin{align}
\frac{\dd r}{\dd k} & = -k^{d+1}\frac{a_d}{d}\frac{u_c(N+2)}{12N}\frac{\omega_{0k}}{\omega_k^4}, \\
\frac{\dd Z_0}{\dd k} & =  Z_0 k^{d+1}\frac{a_d}{d}\frac{u_c(N+2)}{24N}\frac{\omega_{0k}}{\omega_k^6}. \label{eq:Z0-flow}
\end{align}
\end{subequations}
with $a_d = 2/[(4\pi)^{d/2}\Gamma(d/2)]$ a numerical factor ($\Gamma(x)$ is the Gamma function of variable $x$).

The term $\Delta\Gamma_2$ can be evaluated in a similar way as for $\Delta\Gamma_1$ and it contains the renormalization of the vertices $u_c$ and $u_q$. A simple computation (see Ref.~\cite{Chiocchetta2016bis}) renders:
\begin{align}
\label{eq:DGamma2}
\Delta \Gamma_2
& = -\frac{i}{2}\sum_{\alpha\beta \gamma\delta \epsilon}\int_{\rr,\rr',\rr'',t,t',t''}  \text{tr}\left[ G^0_{\alpha\beta}(\rr-\rr',t,t')V_{\beta\gamma}(\rr',t')G^0_{\gamma \delta}(\rr'-\rr'',t',t'')V_{\delta\epsilon}(\rr'',t'')G^0_{\epsilon \alpha}(\rr''-\rr,t'',t) \frac{\dd R}{\dd k}\sigma\right] \nonumber \\
& = -\frac{i}{2}\sum_{\alpha \gamma}\int_{\rr,\rr',\rr'',t,t',t''}  \text{tr}\left[ G^0(\rr-\rr',t,t')V_{\alpha\gamma}(\rr',t')G^0(\rr'-\rr'',t',t'')V_{\gamma\alpha}(\rr'',t'')G^0(\rr''-\rr,t'',t) \frac{\dd R}{\dd k}\sigma\right] \nonumber \\
& \approx -\frac{i}{2}\sum_{\alpha \gamma}\int_{\rr,\rr',\rr'',t,t',t''}  \text{tr}\left[ G^0(\rr-\rr',t,t')V_{\alpha\gamma}(\rr',t')G^0(\rr'-\rr'',t',t'')V_{\gamma\alpha}(\rr',t')G^0(\rr''-\rr,t'',t) \frac{\dd R}{\dd k}\sigma\right] \nonumber \\
& = -\frac{i}{2}\sum_{\alpha \gamma}\int_{\rr,\qq,t,t',t''}  \text{tr}\left[ G^0(q,t,t')V_{\alpha\gamma}(\rr,t')G^0(q,t',t'')V_{\gamma\alpha}(\rr,t')G^0(q,t'',t) \frac{\dd R}{\dd k}\sigma\right] \nonumber \\
& = -ik^{d+1}\frac{a_d}{d}\sum_{\alpha \gamma}\int_{\rr,t,t',t''} \text{tr}\left[ G^0(k,t,t')V_{\alpha\gamma}(\rr,t')G^0(k,t',t'')V_{\gamma\alpha}(\rr,t')G^0(k,t'',t) \sigma\right] \nonumber \\
& = -ik^{d+1}\frac{a_d}{d}\frac{2}{(6N)^2}\sum_{\alpha \gamma}\int_{\rr,t'} 
\biggr\{ 
u_c^2 F_c(t')\left[ 2\phi_c^{(\alpha)}\phi_q^{(\gamma)}  + \delta_{\alpha\gamma}\vecphi_q\cdot\vecphi_c\right]\left[ \delta_{\gamma\alpha}\frac{\vecphi_c^2}{2} +\phi_c^{(\alpha)}\phi_c^{(\gamma)}\right] \nonumber \\
& \qquad \qquad +
u_cu_q F_c(t')\left[ 2\phi_c^{(\alpha)}\phi_q^{(\gamma)}  + \delta_{\alpha\gamma}\vecphi_q\cdot\vecphi_c\right]\left[ \delta_{\gamma\alpha}\frac{\vecphi_q^2}{2} +\phi_q^{(\alpha)}\phi_q^{(\gamma)}\right] \nonumber \\
& \qquad \qquad +
\left[u_c^2 F_{d1}(t')+u_cu_q F_{d2}(t')\right]\left[ \phi_c^{(\alpha)}\phi_q^{(\gamma)} +\phi_c^{(\gamma)}\phi_q^{(\alpha)} + \delta_{\alpha\gamma}\vecphi_q\cdot\vecphi_c\right]\left[ \phi_c^{(\alpha)}\phi_q^{(\gamma)} +\phi_c^{(\gamma)}\phi_q^{(\alpha)} + \delta_{\alpha\gamma}\vecphi_q\cdot\vecphi_c\right]\biggr\}\nonumber \\
& = -ik^{d+1}\frac{a_d}{d}\frac{2}{(6N)^2}\int_{\rr,t'} 
\biggr\{ 
 \frac{N+8}{2}F_c(t')\left[u_c^2 \vecphi_q\cdot\vecphi_c \vecphi_c^2+ u_cu_q \vecphi_q\cdot\vecphi_c \vecphi_q^2\right] \nonumber \\
& \qquad \qquad +
\left[u_c^2 F_{d1}(t')+u_cu_q F_{d2}(t')\right]\left[ 2\vecphi_c^2\vecphi_q^2 + (N+6)(\vecphi_c\cdot\vecphi_q)^2\right]\biggr\}\nonumber \\
\end{align}
with
\begin{align}
F_c(t') &=\int_{0}^{+\infty}\dd t\dd t''\big[ 2G_K(k,t',t'') G_R(k,t,t'')G_R(k,t',t)  + G_K(k,t'',t) G_R(k,t',t)G_R(k,t',t'') \big], \\
F_{d1}(t') & =\int_{0}^{+\infty}\dd t\dd t'' G_K(k,t',t'') G_K(k,t'',t)G_R(k,t',t), \\
F_{d2}(t') & =\int_{0}^{+\infty}\dd t\dd t'' G_R(k,t,t') G_R(k,t'',t)G_R(k,t'',t').
\end{align}
The terms in Eq.~\eqref{eq:DGamma2} proportional to $F_{d1,d2}(t)$ give rise to new operators which were not included in the ansatz~\eqref{eq:ansatz}. However, these operators are irrelevant for what concerns the prethermal fixed point~\cite{Chiocchetta2016}. The renormalization of $u_c$ comes from the term proportional to the function $F_c(t')$, whose explicit expression is given by 
\begin{equation}
F_c(t') = -\frac{\omega_{0k}}{2\omega_k^6} + f_c(t'),
\end{equation}
where $f_c(t')$ is an oscillating function which averages to zero and therefore it is not expected to contribute to the renormalization of $u_c$ at long times. 
%
The  $\beta$-functions for the quantum and classical quartic couplings finally read:
\begin{subequations}
\label{eq:RG-eq-2}
\begin{align}
\frac{\dd u_c}{\dd k} & = \frac{a_d}{d}k^{d+1}\frac{N+8}{6N}\frac{\omega_{0k}}{\omega_k^6} u_c^2, \\
\frac{\dd u_q}{\dd k} & = \frac{a_d}{d}k^{d+1}\frac{N+8}{6N}\frac{\omega_{0k}}{\omega_k^6} u_qu_c.
\end{align}
\end{subequations}

\subsection{Dimensionless $\beta$-functions}
Defining the rescaled couplings $r' \equiv r/k^2 $, $u_{c,q}' \equiv u_{c,q}a_dk^{d-3}/d$, $\widetilde{\Omega}_0 = \Omega_0/k$, the flow equations (20) acquire the dimensionless form
\begin{subequations}
\label{eq:RGquenchSM}
\begin{align}
k\frac{\dd r'}{\dd k} &= -2r' -  \frac{N+2}{12N}\frac{u_c'\sqrt{1+\widetilde{\Omega}_0^2}}{(1+r')^2}, \\
k\frac{\dd u_c'}{\dd k} & = u_c'\left[ d-3 + \frac{N+8}{6N}\frac{u'_c\sqrt{1+\widetilde{\Omega}_0^2}}{(1+r')^3}\right], \\
k\frac{\dd u_q'}{\dd k} & = u_q'\left[ d-3 + \frac{N+8}{6N}\frac{u'_c\sqrt{1+\widetilde{\Omega}_0^2}}{(1+r')^3}\right], \\
k\frac{\dd Z_0}{\dd k} & =  Z_0 \frac{N+2}{24N}  \frac{u_c'\sqrt{1+\widetilde{\Omega}_0^2}}{(1+r')^3},\\
k\frac{\dd \widetilde{\Omega}_0}{\dd k} &= - \widetilde{\Omega}_0.
\end{align}
\end{subequations}
Rewriting Eqs. \eqref{eq:RGquenchSM} in terms of the new couplings $\widetilde{u}_c = u'_c\sqrt{1+\widetilde{\Omega}_0^2}$ and $\widetilde{u}_q = u'_c/\sqrt{1+\widetilde{\Omega}_0^2}$, we find the flow equations reported in Eq.~(4) in the main text.

{\section{Estimate of the thermalization time}}
\label{app:app3}
{In order to provide an estimate of the thermalization time $t_\text{th}$ (or, equivalently, of the duration of the prethermal regime), we consider the generation under RG of a viscosity-like term~\cite{Mitra2011,Mitra2012} in the Keldysh action~\eqref{eq:ansatz}, $\gamma\, \vecphi_q\cdot\dot{\vecphi}_c $, with $\gamma$ a real parameter. This term can be regarded as the inverse of the life-time of quasi-particles and therefore it signals the presence of inelastic processes, which are known to lead to thermalization~\cite{Berges2004b}. While this term is in fact absent in the microscopic theory (i.e., $\gamma=0$ in the initial action), it is generated during the RG flow: the RG-time needed for the running $\gamma$ to become of the same order in magnitude as the other non-irrelevant terms in the Keldysh action heuristically provides an estimate for the thermalization time. In the following, we detail a way to calculate this quantity.}

{We recall that the leading diagrams (in a perturbative expansion in $u$) contributing to the renormalization of $\gamma$ are given by~\cite{Berges2004b}
\begin{equation}
\begin{tikzpicture}[baseline={([yshift=-.5ex]current bounding box.center)}]
\coordinate[] (o) at (0,0);
\coordinate[] (ol) at (-0.5,0);
\coordinate[] (or) at (0.5,0);
\coordinate[] (oll) at (-1.25,0);
\coordinate[] (orr) at (1.25,0);
\coordinate[] (on)  at (0,0.5);
\coordinate[] (on)  at (0,-0.5);
\draw[thick] (or) -- (ol);
\draw[res] (or) arc (0:-90:0.5);
\draw[thick] (or) arc (0:270:0.5);
\draw[res] (ol) node[circle,fill,inner sep=1pt]{} -- (oll);
\draw[thick] (or) node[circle,fill,inner sep=1pt]{} -- (orr);
\end{tikzpicture},
\qquad \text{and} \qquad
\begin{tikzpicture}[baseline={([yshift=-.5ex]current bounding box.center)}]
\coordinate[] (o) at (0,0);
\coordinate[] (ol) at (-0.5,0);
\coordinate[] (or) at (0.5,0);
\coordinate[] (oll) at (-1.25,0);
\coordinate[] (orr) at (1.25,0);
\coordinate[] (on)  at (0,0.5);
\coordinate[] (on)  at (0,-0.5);
\draw[thick] (ol) -- (o);
\draw[res] (o) -- (or);
\draw[res] (or) arc (0:-90:0.5);
\draw[res] (or) arc (0:90:0.5);
\draw[thick] (ol) arc (0:90:-0.5);
\draw[thick] (ol) arc (0:-90:-0.5);
\draw[res] (ol) node[circle,fill,inner sep=1pt]{} -- (oll);
\draw[thick] (or) node[circle,fill,inner sep=1pt]{} -- (orr);
\end{tikzpicture},
\end{equation}
where the normal lines correspond to classical fields, while the wiggled lines correspond to the quantum ones~\cite{Kamenevbook2011}.
Considering, for simplicity, only the first diagram (the second one would only affect the  numerical factor in front of the expression determined by the first), and noticing that the canonical dimension of $\gamma$ at the pre-thermal fixed-point (which can be read straightforwardly from Eq.~\eqref{eq:ansatz}) is $\gamma \sim k$, one may write the following flow equation for the dimensionless parameter $\widetilde{\gamma} = \gamma/k$:
\begin{equation}
\label{eq:gamma-RG}
k \frac{\dd \widetilde{\gamma}}{\dd k} = -\widetilde{\gamma} - \alpha \widetilde{u}_c^2 \frac{N+2}{N^2},
\end{equation}}
{where $\alpha$ is a numerical factor of order unity and $\widetilde{u}_c$ is defined as below Eqs.~\eqref{eq:RGquenchSM}. By introducing the RG flow parameter $\ell = \log(\Lambda/k)$, with $\Lambda$ the UV cutoff of the model, we thus define the thermalization scale $\ell_\text{th}$ from the condition $\widetilde{\gamma}(\ell_\text{th}) \sim 1$. In order to evaluate $\ell_\text{th}$ one needs to solve Eq.~\eqref{eq:gamma-RG} using  $\widetilde{\gamma}(\ell=0) = 0$ as initial condition, which corresponds to a non-dissipative microscopic model.
The solution of Eq.~\eqref{eq:gamma-RG} requires, in principle, to take into account the full dependence of $\widetilde{u}_c$ on $\ell$: however, in order to simplify the calculation, we will assume $\widetilde{u}_c$ to be set at its (quantum or classical) pre-thermal fixed-point value $\widetilde{u}^*_c$. 
One therefore finds
\begin{equation}
\widetilde{\gamma}(\ell) = \alpha (\widetilde{u}^*_c)^2\frac{N+2}{N^2} (\ee^\ell-1),
\end{equation}}
{ which yields, after setting $\ell = \log(\Lambda t)$, the estimate
\begin{equation}
t_\text{th} = \frac{1}{\Lambda} \left[1 + \frac{N^2}{\alpha (\widetilde{u}^*_c)^2 (N+2)}\right].
\end{equation}
For large $N$, it reduces to $t_\text{th} \propto N/(\Lambda (\widetilde{u}^*_c)^2)$ reported in the main text.}

\end{widetext}

\bibliography{biblio}

\end{document}